\documentclass[fleqn,useAMS,usenatbib]{mn2e}

\usepackage{amsmath}
\usepackage{txfonts}
\usepackage{natbib}
\usepackage[dvips]{graphicx}

\usepackage{hyperref}
\usepackage[usenames,dvipsnames]{color}
\definecolor{menublue}{rgb}{0.0,0.0,0.5}
\definecolor{citegreen}{rgb}{0.0,1.0,0.0}
\definecolor{urlred}{rgb}{1.0,0.0,0.0}
\hypersetup{bookmarksopen,pdfstartview={FitH},colorlinks=true, breaklinks=true,menucolor=menublue,urlcolor=urlred,citecolor=citegreen,linkcolor=blue}
\usepackage[all]{hypcap}

\usepackage{graphicx}
\usepackage{caption}
\usepackage{subfigure}

\def\del#1{{}}

\newcommand{\ltsima}{$\; \buildrel < \over \sim \;$}
\newcommand{\lsim}{\lower.5ex\hbox{\ltsima}}
\newcommand{\gtsima}{$\; \buildrel > \over \sim \;$}
\newcommand{\gsim}{\lower.5ex\hbox{\gtsima}}
\newcommand{\bra}{\langle}
\newcommand{\ket}{\rangle}

\newcommand{\dd}{\mathrm{d}}

\newcommand{\ci}{\mathrm{i}}

\newcommand{\trace}{\mathrm{tr}}

\newcommand{\dirac}{\delta_D}

\newcommand{\id}{\mathrm{id}}

\title[angular ellipticity spectra]
{Angular spectra of the intrinsic galaxy ellipticity field, their observability and their impact on lensing in tomographic surveys}
\author[B. M. Sch{\"a}fer and Ph. M. Merkel]
{Bj{\"o}rn Malte Sch\"afer$^{1}$\thanks{e-mail: bjoern.malte.schaefer@uni-heidelberg.de} and Philipp M. Merkel$^2$\\
$^1$Astronomisches Recheninstitut, Zentrum f{\"u}r Astronomie der Universit{\"a}t Heidelberg, Philosophenweg 12, 69120 Heidelberg, Germany\\
$^2$Institut f{\"u}r theoretische Astrophysik, Zentrum f{\"u}r Astronomie der Universit{\"a}t Heidelberg, Philosophenweg 12, 69120 Heidelberg, Germany
}

\begin{document}
\pagerange{\pageref{firstpage}--\pageref{lastpage}}
\pubyear{2015}
\maketitle
\label{firstpage}

\begin{abstract}
Subject of this paper are intrinsic ellipticity correlations between galaxies, their statistical properties, their observability with future surveys and their interference with weak gravitational lensing measurements. Using an angular momentum-based, quadratic intrinsic alignment model we derive correlation functions of the ellipticity components and project them to yield the four non-zero angular ellipticity spectra $C^\epsilon_E(\ell)$, $C^\epsilon_B(\ell)$, $C^\epsilon_C(\ell)$ and $C^\epsilon_S(\ell)$ in their generalisation to tomographic surveys. For a Euclid-like survey, these spectra would have amplitudes smaller than the weak lensing effect on nonlinear structures, but would constitute an important systematic. Computing estimation biases for cosmological parameters derived from an alignment-contaminated survey suggests biases of $+5\sigma_w$ for the dark energy equation of state parameter $w$, $-20\sigma_{\Omega_m}$ for the matter density $\Omega_m$ and $-12\sigma_{\sigma_8}$ for the spectrum normalisation $\sigma_8$. Intrinsic alignments yield a signal which is easily observable with a survey similar to Euclid: While not independent, significances for estimates of each of the four spectra reach values of tens of $\sigma$ if weak lensing and shape noise are considered as noise sources, which suggests relative uncertainties on alignment parameters at the percent level.
\end{abstract}


\section{Introduction}
Intrinsic alignments are a fascinating topic: The idea, that properties and orientation of the luminous stellar component of a galaxy depends on the surrounding dark matter distribution is very persuasive, but the exact mechanisms and parameters are not well understood \citep[for recent reviews, please refer to][]{TI15, Paper1, Paper2, Paper3}. On one side, tidal shearing is thought to be responsible for the alignment of ellipticial galaxies, where the halo potential is distorted by gravitational fields of the large-scale structure. This affects the stellar component as well, leading to a correlation between the brightness distribution and the extrenal tidal gravitational field. On the other side, and this will be the focus of this paper, tidal torquing aims to explain the alignment of spiral galaxies: Here, the stellar disk aligns itself with the halo angular momentum direction, which in turn has been imprinted on the halo by torquing processes during halo formation. This alignment model for spiral galaxies, its observable signatures and its impact on weak lensing is the topic of this paper.

The mechanism of tidal torquing as a perturbative process in structure formation has been quantitatively worked out by \citet{Peebles69} and was first in detail investigated in numerical simulations by \citet{White84, WQS+92, BS05}. Due to the fact that tidal torquing is perturbative \citep{HP88, CT96, Schaefer08}, correlations between angular momenta of galaxies can be traced back to correlations in the tidal fields, giving rise to short-ranged correlations, with typical correlation lengths of about one Mpc \citep{CNP+01, CP01, SM11, LHF+13}, and correlations between the spin-field and density \citep{LP01}. Tidal torquing depends on the environment of the haloes, which can be described by extentions to the standard torquing picture \citep{2015arXiv150406073C}.

Ultimately, tidal torquing leads through the alignments of angular momenta to correlations between the shapes of galaxies, which can be quantified by angular correlation functions or angular spectra, either between the shapes themselves or between the galaxy shape and another property of the large-scale structure. 

Evidence for intrinsic alignments and for tidal torquing in particular has been found in numerical simulations, from simulations with only dark matter \citep{ACC06, LSP+08, FLW+09, AY14} to those with a baryonic component and star formation \citep{2015arXiv150503124T, TMD+14, TMD+14b, schaye15, crain15} aiming to provide a shape of the luminous commponent. Recently, \citet{DPW+14, CGD+14} showed that merging processes play an important role reorienting the angular momentum direction of haloes, which would be in contradiction to the basic idea of predicting angular momentum directions perturbatively.

Observations of intrinsic alignments of spiral galaxies are difficult, and up to now there is not a consistent picture, partially because the large number of different quantifiers employed make comparisons difficult: While early studies have found correlations between density and shapes of galaxies \citep{LP02, MHI+06}, detection reports have been reported for a range of galaxy types in different surveys \citep{BTH+02, LE07, LP07, PSP08, Lee11, HMI+07, SRG09, MBB+11, MSB+13, TL13, TSS13, JSB+13, JSH+13, LJF+13, Zhang13, SMM14}, whereas \citep{AJ11, HvWM+12,HGH+13} did not detect an alignment signal from these galaxies, but instead one due to elliptical galaxies. Therefore, one of the motivation of this work was the derivation of all observable spectra of the intrinsic ellipticity field, and an estimate how well they could potentially be observed with the future Euclid mission\footnote{http://www.euclid-ec.org/} \citep[see][]{LAA+11}.

A second motivation for improving the understanding of intrinsic alignments is weak gravitational lensing: Correlations between galaxy shapes can be due to both effects \citep{LP00, CKB01}. If one were to fit a model for a weak cosmic shear spectrum to data containing both lensing and intrinsic shape correlations, severe parameter estimation biases would be the consequence \citep{KBS10, LBK+12,KRH+12, CMS12, Valageas2014}. This has been investigated in the context of the Dark Energy Survey (DES) and of the weak lensing survey of the Euclid-mission, where large biases have been estimated to occur due to intrinsic alignments. Shape correlations appear on multipoles of $\ell\simeq1000$ for a survey like Euclid, which reaches to redshifts of unity, where the lensing signal is strongest. 

There are a number of ways in which one can deal with intrinsic alignment contamination in weak lensing data, from discarding close pairs of galaxies, to nulling \citep{HutWhi05,JS08}, and ultimately to provide a parameterised model with enough freedom to enable a fit to the combined spectrum with a minimum of assumptions \citep{SB10}. There are already first estimates of the magnitude of this effect, for instance from the Mega-Z survey \citep[see][]{JMA+11}.

At the same time are lensing surveys always intrinsic alignment surveys and can be used to investigate models of galaxy alignment and to determine alignment parameters, if a separation of the two effects is possible \citep{JB10, KBS10}. In fact, gravitational lensing shape correlations and intrinsic alignments differ in their statistical and physical properties, for instance in $B$-mode generation \citep{CNP2002}, in higher-order statistics \citep{SHW+08, SJS10, MS14}, in their appearance in 3d- or tomographic analyses \citep{KS03, MS13} or in their cross-correlations to other data sets \citep{HT14, TI14}. Additionally, there are ways of performing self-calibrations \citep{TI12a,TI12b, Zhang2010a} and downweighting schemes \citep{HH03, King05} that can control intrinsic alignments as a systematic. Conversely, the alignment signal could be enhanced in a statistical way \citep{JS10}.

In summary, we try to answer these questions: Firstly, we aim to derive the full set of intrinsic ellipticity spectra resulting from an angular-momentum based alignment model (Sect.~\ref{sect_ia}), secondly, to quantify their observability in a modern tomographic survey (Sect.~\ref{sect_obs}), and thirdly, to quantify their interference with parameter estimation from weak lensing (Sect.~\ref{sect_lensing}). For the last two points we consider the planned Euclid survey, which reaches to a median redshift of 0.9.

The reference cosmological model used is a spatially flat $w$CDM cosmology with Gaussian adiabatic initial perturbations in the cold dark matter density and a homogeneous dark energy component. The dark energy equation of state-parameter $w$ is constant in time. The specific parameter choices are $\Omega_m = 0.25$, $n_s = 1$, $\sigma_8 = 0.8$, $\Omega_b=0.04$ and $H_0=100\: h\:\mathrm{km}/\mathrm{s}/\mathrm{Mpc}$, with $h=0.7$, and $w=-0.9$. We adopt the summation convention for implied summation over repeated indices.

\section{cosmology}\label{sect_cosmology}

\subsection{Dark energy cosmologies}
In spatially flat dark energy cosmologies with the matter density parameter $\Omega_m$, the Hubble function $H(a)=\dd\ln a/\dd t$ is given by
\begin{equation}
\frac{H^2(a)}{H_0^2} = \frac{\Omega_m}{a^{3}} + (1-\Omega_m)\exp\left(3\int_a^1\dd\ln a\:(1+w(a))\right),
\end{equation}
with the dark energy equation of state $w(a)$. We take $w(a)$ to be a constant with the value $w=-0.9$. The relation between comoving distance $\chi$ and scale factor $a$ is given by
\begin{equation}
\chi = c\int_a^1\frac{\dd a}{a^2 H(a)},
\end{equation}
in units of the Hubble distance $\chi_H=c/H_0$.

\subsection{CDM structures}
The linear CDM density power spectrum $P(k)$ describes the variance of the field $\delta$ in the case of homogeneous Gaussian fluctuations,
$\bra\delta(\bmath{k})\delta(\bmath{k}^\prime)^*\ket=(2\pi)^3\dirac(\bmath{k}-\bmath{k}^\prime)P(k)$, and is given by the inflation-motivated ansatz \citep{BBK+86}
\begin{equation}
P(k)\propto k^{n_s}T^2(k),
\end{equation}
with the transfer function $T(k)$.

The normalisation of the spectrum $P(k)$ is the variance $\sigma_8$ on the scale $R=8~\mathrm{Mpc}/h$,
\begin{equation}
\sigma^2_R = \int\frac{k^2\dd k}{2\pi^2}\: P(k) W^2(kR)
\end{equation}
where a Fourier transformed spherical top hat filter function, $W(x)=3j_1(x)/x$ is used. $j_\ell(x)$ is the spherical Bessel function of the first kind of order $\ell$ \citep{1995mmp..book.....A}.

Growth of the density field, $\delta(\bmath{x},a)=D_+(a)\delta(\bmath{x},a=1)$, in the linear regime $\left|\delta\right|\ll 1$ is described by the growth function $D_+(a)$, which follows as the solution to the growth equation \citep{1997PhRvD..56.4439T, 2003MNRAS.346..573L},
\begin{equation}
\frac{\dd^2 D_+(a)}{\dd a^2} + \frac{2-q(a)}{a}\frac{\dd D_+(a)}{\dd a} =
\frac{3\Omega_m(a)}{2a^2} D_+(a),
\label{eqn_growth}
\end{equation}
where both $\Omega_m(a)$ and the deceleration parameter $q(a)$ carry the dependence on dark energy.

The gravitational potential $\Phi$ (normalised by $c^2$), which we use to predict the two observables gravitational lensing shear and galaxy ellipticity, is related to the growing density field through the comoving Poisson equation,
\begin{equation}
\Delta\Phi = \frac{3\Omega_m}{2\chi_H^2}\frac{\delta}{a},
\end{equation}
such that the gravitational potential grows $\propto D_+(a)/a$. For gravitational lensing on nonlinear scales we use the model by \citet{2003MNRAS.341.1311S}. In contrast, the intrinsic alignment spectra are computed from a linear CDM-spectrum, because tidal torquing is a perturbative process and is consistently computed from the initial conditions of structure formation.

\subsection{Weak gravitational lensing}
Weak lensing refers to the correlated change in the shapes of distant galaxies by gravitational deflection and distortion by potentials in the cosmic large-scale structure \citep[for reviews, see][]{Ba10,BS01}, which is one of the prime ways to measure cosmological parameters \citep[summarised in][]{2008ARNPS..58...99H, K14}. The two components $\gamma_{+,i}$ and $\gamma_{\times,i}$ of the complex weak lensing shear $\gamma_i=\gamma_{+,i}+\ci\gamma_{\times,i}$ in tomography bin $i$ with edges at comoving distances $\chi_i$ and $\chi_{i+1}$ provide a mapping of the projected second derivatives of the gravitational potential $\Phi$ \citep{TJ04,H99,TW04},
\begin{eqnarray}
\gamma_{+,i} & = & \int_0^{\chi_H}\dd\chi\:W_i(\chi)\left(\partial^2_y-\partial^2_x\right)\Phi\\
\gamma_{\times,i} & = & \int_0^{\chi_H}\dd\chi\:W_i(\chi)\left(2\:\partial^2_{xy}\right)\Phi
\end{eqnarray}
weighted by the lensing efficiency function
\begin{equation}
W_i(\chi) = \frac{D_+(a)}{a}\:G_i(\chi)\chi
\end{equation}
with
\begin{equation}
G_i(\chi) = \int_{\mathrm{max}(\chi,\chi_i)}^{\chi_{i+1}}\dd\chi^\prime n(\chi^\prime)\:\frac{\chi^\prime-\chi}{\chi^\prime}.
\end{equation}
Here, $n(\chi)=n(z)\dd z/\dd\chi = n(z)H(z)$ is the distance distribution of the galaxy sample and $\chi_i$ denote the bin boundaries in terms of comoving distance.

Correlation properties of the two shear components can be described by two correlation functions \citep{K92},
\begin{eqnarray}
C^\gamma_{+,ij}(\theta) & = & \bra\gamma_{+,i}\gamma_{+,j}^\prime\ket + \bra\gamma_{\times,i}\gamma_{\times,j}^\prime\ket,\\
C^\gamma_{-,ij}(\theta) & = & \bra\gamma_{+,i}\gamma_{+,j}^\prime\ket - \bra\gamma_{\times,i}\gamma_{\times,j}^\prime\ket.
\end{eqnarray}
Subsequent Fourier-transformation yields the two tomographic spectra $C^\gamma_{E,ij}(\ell)$ and $C^\gamma_{B,ij}(\ell)$ \citep{2002A&A...389..729S, 2007A&A...462..841S, 2010MNRAS.401.1264F},
\begin{eqnarray}
C^\gamma_{E,ij}(\ell) & = & \pi\int\theta\dd\theta\:\left(C^\gamma_{+,ij}(\theta)J_0(\ell\theta) + C^\gamma_{-,ij}(\theta)J_4(\ell\theta)\right),\\
C^\gamma_{B,ij}(\ell) & = & \pi\int\theta\dd\theta\:\left(C^\gamma_{+,ij}(\theta)J_0(\ell\theta) - C^\gamma_{-,ij}(\theta)J_4(\ell\theta)\right).
\end{eqnarray}

The positive-parity spectrum $C^\gamma_{E,ij}(\ell)$ is the primary lensing observable, while the negative-parity spectrum $C^\gamma_{B,ij}(\ell)$ is zero to lowest order in lensing, in contrast to intrinsic alignments, where both ellipticity spectra $C^\epsilon_{E,i}(\ell)$ and $C^\epsilon_{B,i}(\ell)$ are nonzero, and in fact measurable.

For everything related to weak lensing, we use the specification of a Euclid-like weak lensing survey: median redshift $z_\mathrm{med}=0.9$, density of 40 lensed galaxies per squared arcminute and an ellipticity shape noise of $\sigma_\epsilon=0.3$. The specific redshift distribution $n(z)\dd z$ is parameterised by:
\begin{equation}
n(z)\dd z = n_0\left(\frac{z}{z_0}\right)^2\exp\left(-\left(\frac{z}{z_0}\right)^\beta\right)\dd z
\quad\mathrm{with}\quad
\frac{1}{n_0}=\frac{z_0}{\beta}\Gamma\left(\frac{3}{\beta}\right).
\end{equation}
with the parameters $\beta=3/2$ and $z_0=z_\mathrm{med}/\sqrt{2}$, and tomographic redshift bins are chosen to contain an equal fraction of the total number of galaxies. Likewise, the bin boundaries $\chi_i$ for weak lensing tomography, and later for the ellipticity correlations, are chosen to contain equal fractions of the total galaxies, which keeps the shape noise in each bin constant.

\section{Intrinsic alignments}\label{sect_ia}

\subsection{Angular momenta of galaxies}
The physical picture for angular-momentum induced alignments in spiral galaxies is the following: The host CDM halo picks up angular momentum through tidal torquing \citep{Peebles69,White84} prior to gravitational collapse. In this process, both baryonic and dark matter have equal amounts of specific angular momentum, and when at a much later stage a galactic disk is formed, its symmetry axis should be aligned with the host halo angular momentum axis, if neither dissipative processes nor merging activity have have changed the angular momentum or the disk orientation. Commonly, one uses the alignment model by \citet{LP00,LP01} for setting up a distribution $p(\hat{L}|\hat{\Phi}_{\alpha\beta})\dd\hat{L}$ for the angular momentum direction $\hat{L}$ conditional on the tidal shear field $\partial^2_{\alpha\beta}\Phi$ through the covariance matrix,
\begin{equation}
\bra \hat{L}_\alpha\hat{L}_\beta\ket = 
\frac{1}{3}\left(\frac{1+a}{3}\delta_{\alpha\beta}-a\:\hat{\Phi}_{\alpha\gamma}\hat{\Phi}_{\gamma\beta}\right)
\end{equation}
where $\hat{\Phi}_{\alpha\beta}$ is the unit-normalised traceless tidal shear. The parameter $a$ interpolates between isotropic and tightly coupled angular momentum directions and is measured to be $a\simeq1/4$ in numerical simulations for galaxy-sized haloes. It effectively describes the amount of misalignment between the gravitational shear tensor and the moment of inertia of the halo prior to collapse \citep{PDH02a,PDH02b}. We emphasise that our model links angular momentum and ultimately galaxy shape to the squared tidal shear in an averaged way without distinguishing between environments whose dynamical properties can have an influence on the spin statistics \citep[see, for instance,][]{KO92,CBG+08,AWJ+07,2015arXiv150406073C}. Assuming perturbation theory allows to link the angular momentum direction to the initial conditions of structure formation: This assumption has been tested by \citet{PDH02a,PDH02b,LP08} in simulations, who found that the angular momentum direction can be predicted well, in contrast to the angular momentum magnitude.

\subsection{Angular momentum alignment}
The idea of angular-momentum induced alignments (so-called quadratic alignments) is an alignment of the symmetry axis of the galactic disk with the host halo angular momentum direction \citep{HRH2000, CNP+01, CNP2002, MWK02}. The ellipticity $\epsilon=\epsilon_++\ci\epsilon_\times$ measured by an observer then depends on the direction $\hat{L}=\bmath{L}/L$ of the angular momentum vector $\bmath{L}$ relative to the line of sight:
\begin{equation}
\epsilon_+ = \alpha\frac{\hat{L}_y^2-\hat{L}_x^2}{1+\hat{L}_z^2}
\quad\mathrm{and}\quad
\epsilon_\times = 2\alpha\frac{\hat{L}_x\hat{L_y}}{1+\hat{L}_z^2},
\end{equation}
if the line of sight is taken to be the $z$-direction. The modulus $\left|\epsilon\right|$ of the ellipticity depends on $\hat{L}$ according to
\begin{equation}
\left|\epsilon\right| = \alpha\frac{1-\hat{L}_z^2}{1+\hat{L}_z^2},
\end{equation}
by using the normalisation $\hat{L}^2=1$ of the angular momentum direction. $\alpha$ is a phenomenological parameter which weakens the dependence of $\epsilon$ on $\hat{L}$ if taken to be smaller than one, and in this work we adopt the conservative value of $\alpha=1/2$. Pictorially, the parameter describes the effect of a galactic disk of finite thickness, where the change in ellipticity with inclination angle is weaker compared to an idealised infinitely thin disk. \citet{CNP+01} have computed the mean value of the complex ellipticity $\epsilon$ for a given tidal shear averaged over angular momentum magnitudes and directions, and arrive at
\begin{equation}
\epsilon(\hat{\Phi}) = \frac{a\alpha}{2}\sum_\alpha\left(\hat{\Phi}_{x\alpha}\hat{\Phi}_{\alpha x}-\hat{\Phi}_{y\alpha}\hat{\Phi}_{\alpha y}-2\ci\hat{\Phi}_{x\alpha}\hat{\Phi}_{\alpha y}\right).
\label{eqn_complex}
\end{equation}
Similarly, the absolute value $\epsilon_s=\left|\epsilon\right|=\sqrt{\epsilon_+^2+\epsilon_\times^2}$ of the complex ellipticity $\epsilon=\epsilon_++\ci\epsilon_\times$ can be related to the tidal shear through
\begin{equation}
\epsilon_s = \frac{3}{4}a\alpha\sum_\alpha\hat{\Phi}_{z\alpha}\hat{\Phi}_{\alpha z},
\label{eqn_scalar}
\end{equation}
using the normalisation condition $\hat{\Phi}_{\alpha\beta}\hat{\Phi}_{\beta\alpha}=1$. It should be emphasised that the scalar ellipticity $\epsilon_s$ is not statistically independent from the complex ellipticity $\epsilon=\epsilon_++\ci\epsilon_\times$ and that it will in fact contain less information than the complex ellipticity due to removal of the ellipse's phase angle, i.e. its orientation, while only keeping information about the axis ratio. This type of information is as well a source of lensing information as shown by \citet{HAJ13, AKH+14}

\subsection{Ellipticity correlations}
Neighbouring galaxies have formed from similar initial conditions and show ellipticity correlations through correlations in the tidal shear they have been exposed to while building up their angular momenta. It is important to emphasise that alignment processes are thought to be local and do not arise through interaction between the galaxies: The are generated only through correlations in the aligning tidal field $\partial^2_{\alpha\beta}\Phi$. Therefore, correlations in $\epsilon$ must be expressed in terms of correlations of squares of the tidal shear $\hat{\Phi}$. This proceeds in two steps. Firstly, the tidal shear field correlations $C_{\alpha\beta\gamma\delta}(r)\equiv\bra\Phi_{\alpha\beta}(\bmath{x})\Phi_{\gamma\delta}(\bmath{x}^\prime)\ket$, $r=\left|\bmath{x}-\bmath{x}^\prime\right|$, can be computed to be
\begin{equation}
\begin{split}
C_{\alpha\beta\gamma\delta}(r) = &
(\delta_{\alpha\beta}\delta_{\gamma\delta}+\delta_{\alpha\gamma}\delta_{\beta\delta}+\delta_{\alpha\delta}\delta_{\beta,\gamma})\zeta_2(r)+\\ 
&(r_\alpha r_\beta \delta_{\gamma\delta}+\mathrm{5~perm.}) \zeta_3(r)+\\
&r_\alpha r_\beta r_\gamma r_\delta \zeta_4(r)
\end{split}
\end{equation}
from which the correlations $\tilde{C}_{\alpha\beta\gamma\delta}(r)$ of the traceless shear follow by subtraction of the trace $\Phi_{\alpha\alpha}/3\:\delta_{\alpha\beta}$ from the tidal shear $\Phi_{\alpha\beta}$. The unit vector $r_\alpha$ is chosen to have the entries $(\sin\alpha,0,\cos\alpha)$ by suitable rotation of the coordinate frame. Finally, correlations of the squared tidal shear components are related to $\tilde{C}_{\alpha\beta\gamma\delta}(r)$ by virtue of
\begin{equation}
\bra\tilde{\Phi}_A(\bmath{x})\tilde{\Phi}_B(\bmath{x})\:\tilde{\Phi}_C(\bmath{x}^\prime)\tilde{\Phi}^\prime_D(\bmath{x}^\prime)\ket = 
\frac{1}{(14\zeta_2(0))^2}\left(\tilde{C}_{AC}\tilde{C}_{BD}+\tilde{C}_{AD}\tilde{C}_{BC}\right),
\end{equation}
where the capital letters denote the index pairs of the tidal shear tensors $\Phi_{\alpha\beta}(\bmath{x})$. 

We generalise the derivation presented by \citet{CNP+01} to include all possible correlations between the ellipticity components $\epsilon_+$, $\epsilon_\times$ and $\epsilon_s$. They follow from contraction of the expressions~(\ref{eqn_complex}) and~(\ref{eqn_scalar}) while substituting $r_\alpha$ for our orientation of the coordinate frame and while making use of the Kronecker-$\delta$ symbol.
\begin{eqnarray}
\bra\epsilon_+\epsilon_+^\prime\ket & = &
\frac{1}{144}\left(\frac{a\alpha}{14\zeta_2}\right)^2
\left(
A_{++}\cos(4\alpha) +
B_{++}\cos(2\alpha) +
C_{++}
\right)\\
\bra\epsilon_\times\epsilon_\times^\prime\ket & = &
\frac{1}{18}\left(\frac{a\alpha}{14\zeta_2}\right)^2
\left(
B_{\times\times} \cos(2\alpha) +
C_{\times\times}
\right)\\
\bra\epsilon_+\epsilon_s\ket & = &
\frac{1}{324}\left(\frac{a\alpha}{14\zeta_2}\right)^2
\left(
A_{+s}\cos(4\alpha) +
B_{+s}\cos(2\alpha) +
C_{+s}
\right)\\
\bra\epsilon_s\epsilon_s\ket & = &
\frac{1}{108}\left(\frac{a\alpha}{14\zeta_2}\right)^2
\left(
A_{ss}\cos(4\alpha) +
B_{ss}\cos(2\alpha) + 
C_{ss}
\right)
\end{eqnarray}
with the abbreviations
\begin{eqnarray}
A_{++} & = &\zeta^2_4+6(\zeta_3+\zeta_2)\zeta_4+17\zeta^2_3\\
B_{++} & = &-4\zeta^2_4-32\zeta_3\zeta_4-28\zeta^2_3+72\zeta_2\zeta_3\\
C_{++} & = &3\zeta^2_4+(26\zeta_3+58\zeta_2)\zeta_4+155\zeta^2_3+472\zeta_2\zeta_3+336\zeta_2^2
\end{eqnarray}
for the correlation function $\bra\epsilon_+\epsilon_+^\prime\ket$,
\begin{eqnarray}
A_{\times\times} & = & 0\\
B_{\times\times} & = &(\zeta_3-3\zeta_2)\zeta_4-5\zeta^2_3-9\zeta_2\zeta_3\\
C_{\times\times} & = &(-\zeta_3-5\zeta_2)\zeta_4-13\zeta^2_3-59\zeta_2\zeta_3-42\zeta_2^2
\end{eqnarray}
for the function $\bra\epsilon_\times\epsilon_\times^\prime\ket$ with $A_{\times\times}$ being zero. These results have been derived by \citet{CNP+01}, and are listed for completeness. In addition, we supply the new results
\begin{eqnarray}
A_{+s} & = &-3\zeta^2_4+18(\zeta_3+\zeta_2)\zeta_4+51\zeta^2_3\\
B_{+s} & = &4\zeta^2_4+(32\zeta_3+48\zeta_2)\zeta_4+172\zeta^2_3+408\zeta_2\zeta_3\\
C_{+s} & = &7\zeta^2_4+(-50\zeta_3-66\zeta_2)\zeta_4-233\zeta_3^2-408\zeta_2\zeta_3
\end{eqnarray}
for the correlation $\bra\epsilon_+\epsilon_s^\prime\ket$ and lastly
\begin{eqnarray}
A_{ss} & = &9 \zeta^2_4+(54 \zeta_3+54 \zeta_2)\zeta_4+153 \zeta^2_3\\
B_{ss} & = &60 \zeta^2_4+(480 \zeta_3+288 \zeta_2)\zeta_4+1284 \zeta^2_3+1800 \zeta_2\zeta_3\\ 
C_{ss} & = &59 \zeta^2_4+(490 \zeta_3 +426 \zeta_2)\zeta_4 +1907 \zeta^2_3 +3864 \zeta_2 \zeta_3 +2448 \zeta^2_2\nonumber
\end{eqnarray}
for $\bra\epsilon_s\epsilon_s^\prime\ket$. We checked that the remaining correlation functions $\bra\epsilon_+\epsilon_\times\ket$ and $\bra\epsilon_\times\epsilon_s\ket$ are in fact zero. The function $\zeta_n(r)$,
\begin{equation}
\zeta_n(r) = \frac{(-1)^n}{r^{4-n}}\int\frac{\dd k}{2\pi^2}\:k^{n-2}j_n(kr)P(k)
\end{equation}
denotes weighted moments of the CDM-spectrum $P(k)$. Here, we impose a Gaussian smoothing on the CDM-spectrum $P(k)$,
\begin{equation}
P(k) \rightarrow P(k)\exp\left(-\frac{(kR)^2}{2}\right),
\end{equation}
where the smoothing scale $R$ is linked to the mass scale of the objects under consideration through $M=4\pi/3\:\Omega_m\rho_\mathrm{crit}R^3$. Specifically, we set the mass scale to $M=10^{12}M_\odot/h$, which is typical for a spiral galaxy. Because spin generation in a halo is taking place prior to gravitational collapse, the smoothing scale uses the cosmic value $\Omega_m\rho_\mathrm{crit}$ for the density. We work with a Gaussian smoothing for numerical purposes.

In the very last step, Limber projection of the 3d-correlation functions $\bra\epsilon\epsilon^\prime\ket(r,\alpha)$ with the galaxy distribution $n(\chi)$ in the corresponding tomography bin $i$ yields the angular ellipticity correlation functions $\bra\epsilon\epsilon^\prime\ket(\theta)$ for all four non-zero combinations of ellipticity modes. The angular correlation functions depend on a single bin index, i.e. there are no cross correlation of the ellipticity field between different bins, because the correlation length of the ellipticity field is much shorter than the bin width: Typical numbers would be a few Mpc for the ellipticity field in comparison to hundreds of Mpc for Euclid's tomography bin size.

\subsection{Ellipticity correlation function}
The two components of ellipticity can be combined into a complex ellipticity $\epsilon = \epsilon_++\ci\epsilon_\times$, such that in analogy to weak lensing or to the polarisation of the cosmic microwave background two correlation functions $C^\epsilon_{+,i}(\theta)$ and $C^\epsilon_{-,i}(\theta)$ can be defined:
\begin{eqnarray}
C^\epsilon_{+,i}(\theta) & = & \bra\epsilon_{+,i}\epsilon_{+,i}^\prime\ket + \bra\epsilon_{\times,i}\epsilon_{\times,i}^\prime\ket,\\
C^\epsilon_{-,i}(\theta) & = & \bra\epsilon_{+,i}\epsilon_{+,i}^\prime\ket - \bra\epsilon_{\times,i}\epsilon_{\times,i}^\prime\ket.
\end{eqnarray}
These correlation functions can be complemented by those involving the modulus $\epsilon_s$ of the ellipticity, a cross correlation $C^\epsilon_{C,i}(\theta)$ and an auto-correlation $C^\epsilon_{S,i}(\theta)$:
\begin{eqnarray}
C^\epsilon_{C,i}(\theta) & = & \bra\epsilon_{+,i}\:\epsilon_{s,i}^\prime\ket,\\
C^\epsilon_{S,i}(\theta) & = & \bra\epsilon_{s,i}\:\epsilon_{s,i}^\prime\ket.
\end{eqnarray}
In contrast to lensing, in which correlations between different tomography bins are natural because the light from galaxies inside a distant bin needs to transverse foreground bins which gives rise to a correlated lensing signal, the ellipticity correlations only occur locally, due to the short-rangedness of tidal fields.

\begin{figure}
\begin{center}
\resizebox{\hsize}{!}{\includegraphics{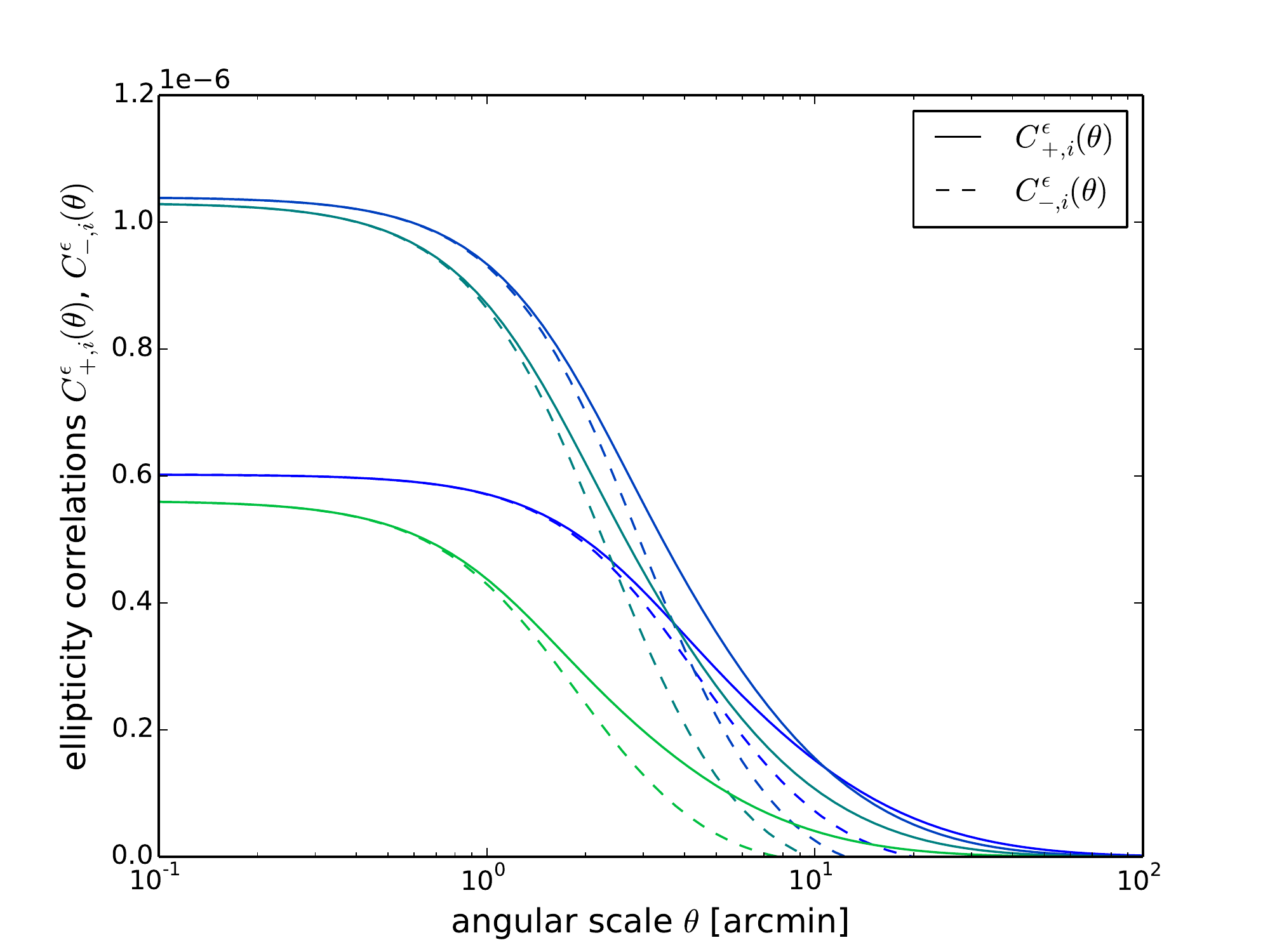}}
\end{center}
\caption{Ellipticity correlation functions $C^\epsilon_{+,i}(\theta)$ (solid lines) and $C^\epsilon_{-,i}(\theta)$ (dashed lines) for 4-bin-tomography with a Euclid-like survey.}
\label{fig_angular_eb}
\end{figure}

Fig.~\ref{fig_angular_eb} shows the two correlation functions $C^\epsilon_{+,i}(\theta)$ and $C^\epsilon_{-,i}(\theta)$ for a 4-bin tomographic survey. Clearly, intrinsic ellipticity correlations are short-ranged and only present on scales below a few arcminutes. In both cases the ellipticity field shows correlations of the order $10^{-6}$ on small scales, which is smaller than that caused by gravitational lensing. Only on very small scales the two effects will become comparable.

\begin{figure}
\begin{center}
\resizebox{\hsize}{!}{\includegraphics{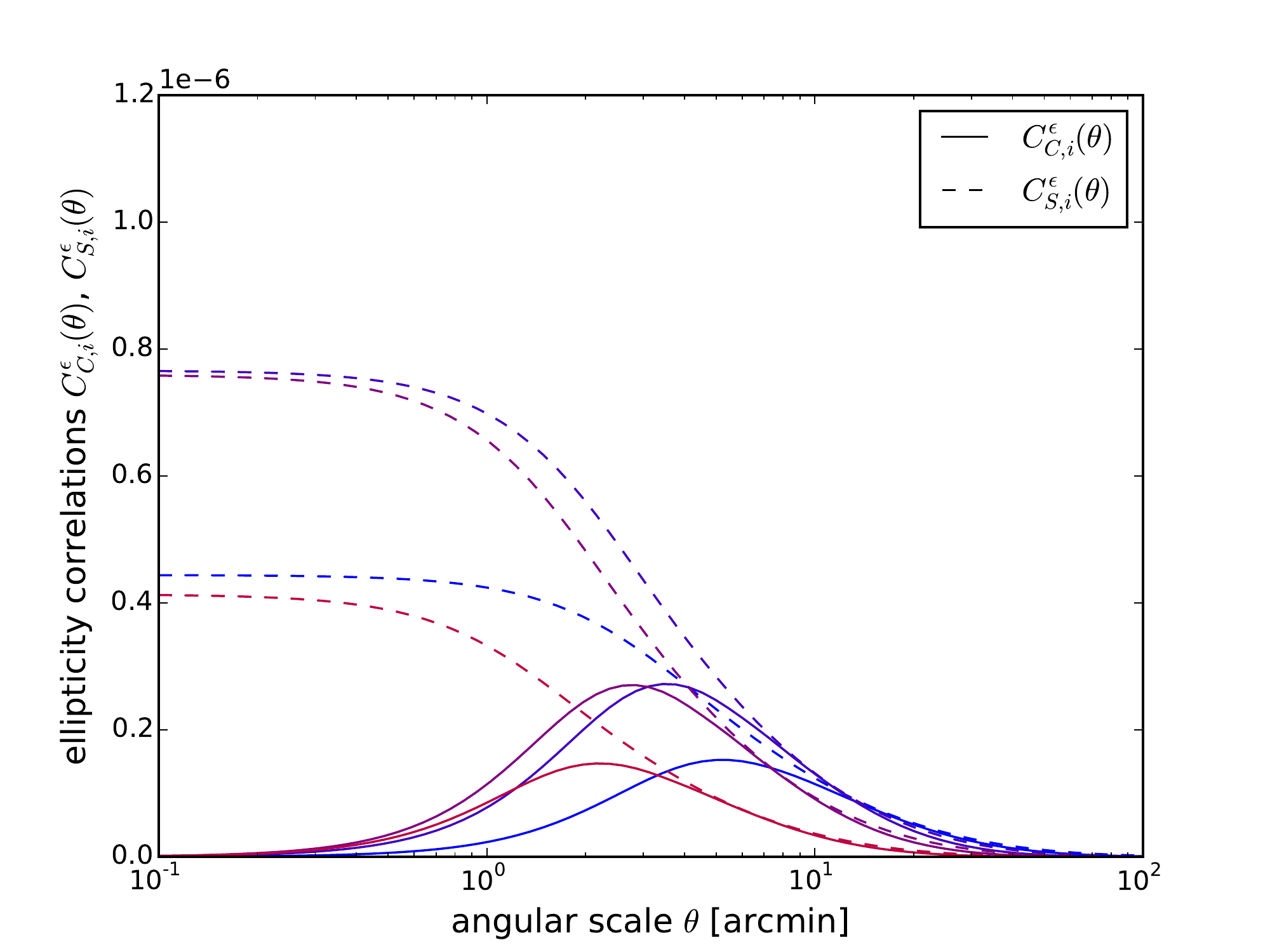}}
\end{center}
\caption{Ellipticity correlation functions $C^\epsilon_{C,i}(\theta)$ (solid lines) and $C^\epsilon_{S,i}(\theta)$ (dashed line) for 4-bin tomography with a Euclid-like survey. The scaling has been chosen to make this figure directly comparable to Fig.~\ref{fig_angular_eb}.}
\label{fig_angular_sc}
\end{figure}

The correlation functions $C^\epsilon_{C,i}(\theta)$ and $C^\epsilon_{S,i}(\theta)$ are depicted in Fig.~\ref{fig_angular_sc}. Again, sizeable amplitudes are only present below the scale of a few arcminutes, and fall short in comparison to those of gravitational lensing.

\subsection{Angular ellipticity spectra}
In complete analogy to weak lensing, in total four nonzero ellipticity spectra can be derived from the ellipticity correlation functions. The first pair is
\begin{eqnarray}
C^\epsilon_{E,i}(\ell) & = & \pi\int\theta\dd\theta\:\left(C^\epsilon_{+,i}(\theta)J_0(\ell\theta) + C^\epsilon_{-,i}(\theta)J_4(\ell\theta)\right),\\
C^\epsilon_{B,i}(\ell) & = & \pi\int\theta\dd\theta\:\left(C^\epsilon_{+,i}(\theta)J_0(\ell\theta) - C^\epsilon_{-,i}(\theta)J_4(\ell\theta)\right),
\end{eqnarray}
from the correlation functions $C^\epsilon_{+,i}(\theta)$ and $C^\epsilon_{-,i}(\theta)$ and the second pair
\begin{eqnarray}
C^\epsilon_{C,i}(\ell) & = & 2\pi\int\theta\dd\theta\:C^\epsilon_{C,i}(\theta)J_2(\ell\theta),\\
C^\epsilon_{S,i}(\ell) & = & 2\pi\int\theta\dd\theta\:C^\epsilon_{S,i}(\theta)J_0(\ell\theta),
\end{eqnarray}
while all other combinations vanish due to statistical parity invariance.

\begin{figure}
\begin{center}
\resizebox{\hsize}{!}{\includegraphics{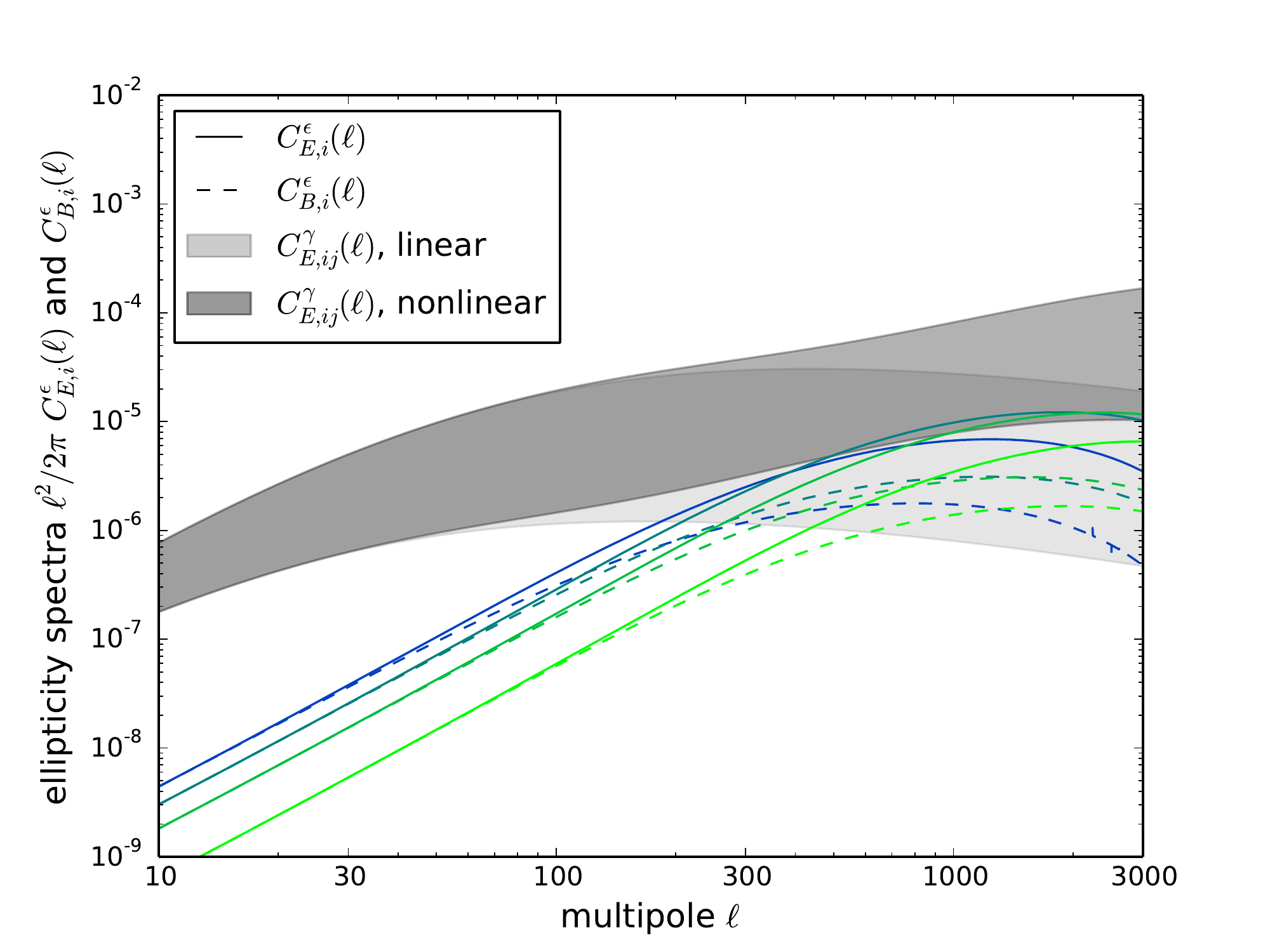}}
\end{center}
\caption{Angular ellipticity spectra $C^\epsilon_{E,i}(\ell)$ (solid lines) and $C^\epsilon_{B,i}(\ell)$ (dashed lines) in comparison to weak lensing spectra $C^\gamma_{E,ij}(\ell)$ (shaded area) for linear (light grey) and nonlinear CDM-spectra (dark grey), for 4 tomography bins. The colour gradient ranges from blue for the low-redshift bins to green for the high-redshift bins.}
\label{fig_spectrum_eb}
\end{figure}

The spectra $C^\epsilon_{E,i}(\ell)$ and $C^\epsilon_{B,i}(\ell)$ are shown in Fig.~\ref{fig_spectrum_eb} alongside a grey band which shows the amplitude of the weak lensing spectrum $C^\gamma_{E,ij}(\ell)$ from linear and from nonlinear structures, computed for a 4-bin survey. At low multipoles both spectra are constant and equally large, only at multipoles above a few hundred the spectrum $C^\epsilon_E(\ell)$ dominates over the spectrum $C^\epsilon_B(\ell)$ by about an order of magnitude. The shape of the spectrum depends on the redshift interval of the tomography bin because correlations of the same physical correlation length appear through the choice of redshift bin on different angular scales. Ultimately, the spectra drop of rapidly due to the exponential smoothing of the CDM-spectrum. It is interesting to not that, with our conservative choice for the misalignment parameter $a$ and the disk thickness $\alpha$ intrinsic alignments provide a significant contribution to the $E$-mode weak lensing spectrum, which will be the focus of Sect.~\ref{sect_lensing}.

\begin{figure}
\begin{center}
\resizebox{\hsize}{!}{\includegraphics{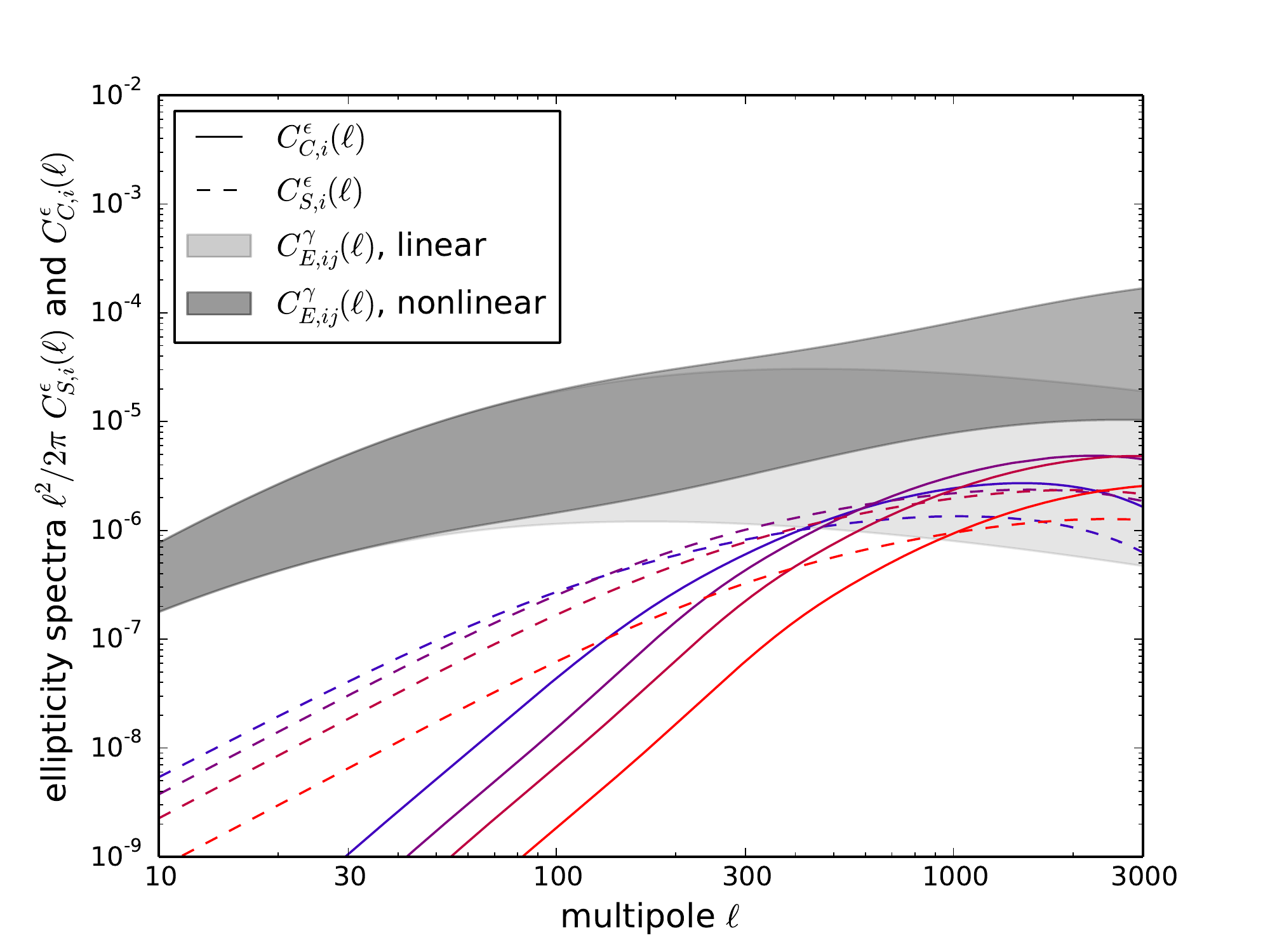}}
\end{center}
\caption{Angular ellipticity spectra $C^\epsilon_{C,i}(\ell)$ (solid lines) and $C^\epsilon_{S,i}(\ell)$ (dashed lines) in comparison to weak lensing spectra $C^\gamma_{E,ij}(\ell)$ (shaded area) for linear (light grey) and nonlinear structure formation (dark grey) and for 4 tomography bins. The colour gradient changes from blue for the low-redshift bins to red for the high-redshift bins.}
\label{fig_spectrum_sc}
\end{figure}

Finally, the spectra $C^\epsilon_{C,i}(\ell)$ and $C^\epsilon_{S,i}(\ell)$ are given in Fig.~\ref{fig_spectrum_sc}, again in comparison to the weak lensing spectrum $C^\gamma_{E,ij}(\ell)$ for a 4-bin survey. They tend to have smaller amplitudes compared to $C^\epsilon_{E,i}(\ell)$ or $C^\epsilon_{B,i}(\ell)$, and in particular $C^\epsilon_{C,i}(\ell)$ has much smaller amplitudes at low multipoles. 

\section{Observability}\label{sect_obs}
A natural question concerns the significance at which intrinsic alignments can be observed in future, possibly tomographic surveys. For this purpose, we compute a forecast for the signal to noise-ratio which can be reached with a Euclid-like survey under the assumption of Gaussian statistics and a perfect separation between weak lensing and intrinsic alignment correlation. Noise sources considered are shape noise in the ellipticity measurement and cosmic variance. In a previous paper \citep{CMS12} we have shown at least for a non-tomographic survey, that using priors on cosmological parameters from e.g. observations of the cosmic microwave background and from baryon acoustic oscillations allow a sufficiently accurate prediction of the weak lensing contribution of the spectrum such that it can be subtracted.

We estimate the signal to noise-ratio using the Fisher-formalism for determining the error on the amplitude of the signal and compute the signal to noise-ratio as the inverse relative error on the unknown normalisation of the signal. Specifically, the signal strength $\Sigma$ follows then from
\begin{equation}
\Sigma^2 = \frac{f_\mathrm{sky}}{2}\sum_\ell(2\ell+1)\:\trace\left(C^{-1}S\:C^{-1}S\right)
\end{equation}
with the signal covariance $S_{ij}$ and the total covariance $C_{ij}$, which contains in addition to the intrinsic ellipticity spectra the contribution from weak gravitational lensing and ellipticity shape noise. In this way the cosmic variance, which is mostly generated by gravitational lensing, is properly taken care of. The sum is extended over all multipoles up to $\ell=3000$ to capture most of the signal before the shape noise contribtion dominates.

\begin{figure}
\begin{center}
\resizebox{\hsize}{!}{\includegraphics{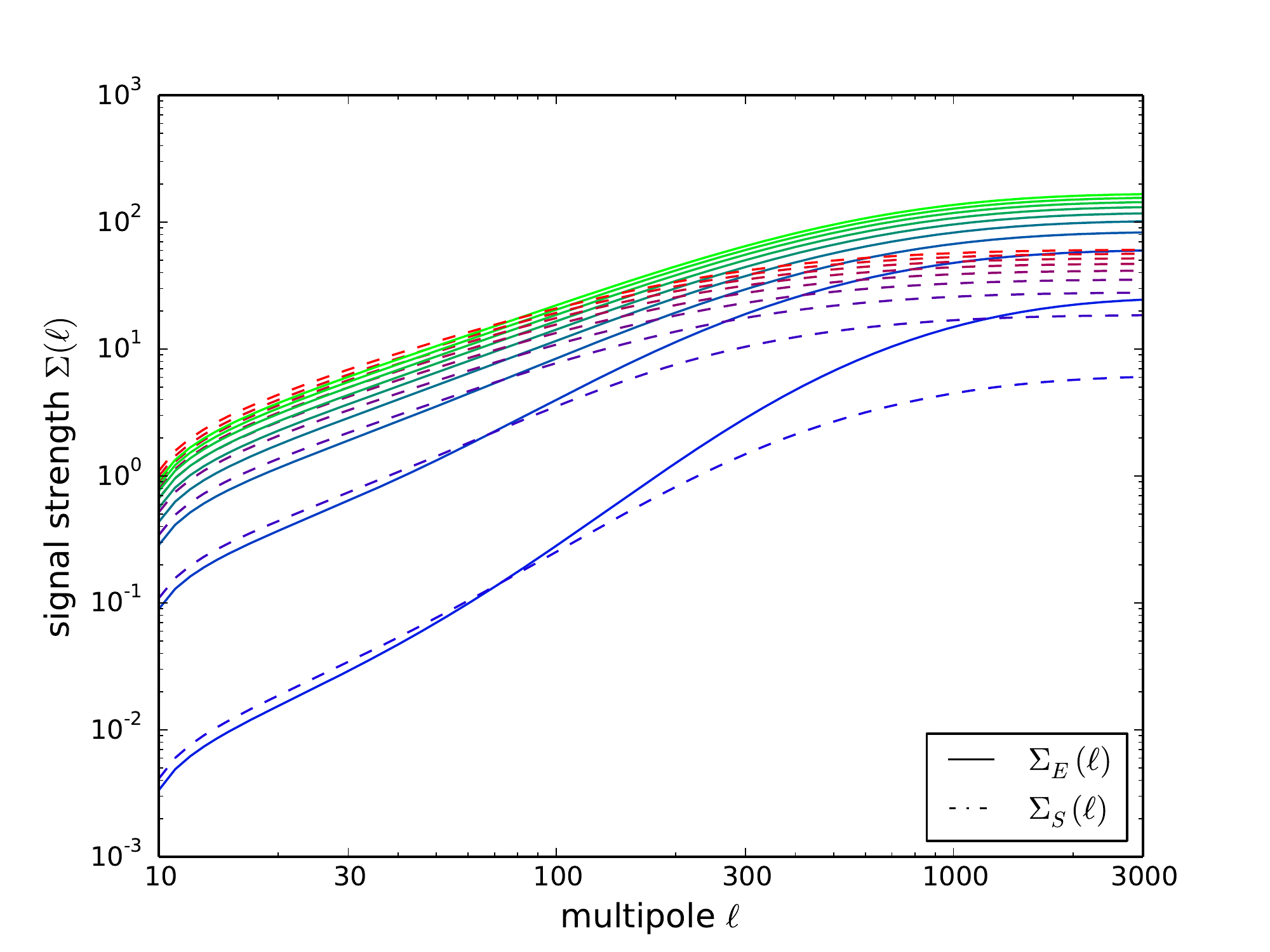}}
\end{center}
\caption{Signal amplitude for observing the ellipticity spectra $C^\epsilon_{E,i}(\ell)$ (solid lines) and $C^\epsilon_{S,i}(\ell)$ (dashed lines) as a function of multipole $\ell$ and the total number of redshift bins into which the survey is divided. The survey was assumed to cover the full sky. The signal amplitude increases with the number of tomography bins.}
\label{fig_s2n_es}
\end{figure}

Specifically, we consider the case of separately measuring the positive parity spectra $C^\epsilon_{E,i}(\ell)$ and $C^\epsilon_{S,i}(\ell)$ first. As shown by Fig.~\ref{fig_s2n_es}, the signal increases with increasing multipole $\ell$ until it levels off at $\ell\simeq1000$ when the shape noise of the ellipticity becomes dominant. Subdivision of the galaxy samples in tomographic bins is able to boost the signal significantly from $\Sigma_S = 6\sigma$ ($3.4\sigma$) to $\Sigma_S = 60\sigma$ ($34\sigma$) in the case of $C^\epsilon_S(\ell)$ and from $\Sigma_E = 25\sigma$ ($14\sigma$) to close to $\Sigma_E = 200\sigma$ ($114\sigma$) in the case of $C^\epsilon_E(\ell)$. The number in parantheses correspond to Euclid's sky coverage of $f_\mathrm{sky}=0.33$, which reduce the full sky significances by about 30\%, and we considered tomography with up to 9 bins in redshift.

\begin{figure}
\begin{center}
\resizebox{\hsize}{!}{\includegraphics{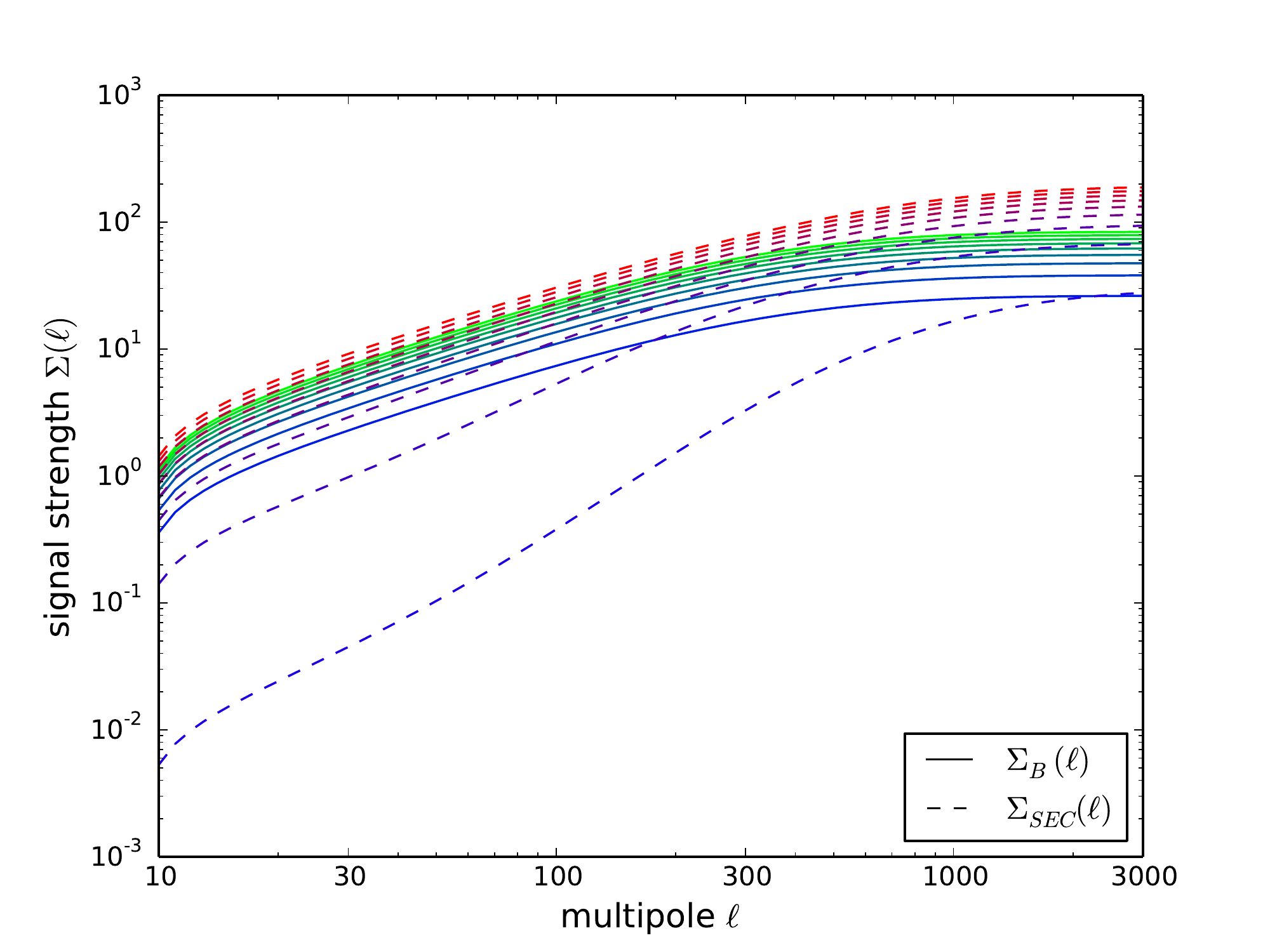}}
\end{center}
\caption{Signal amplitude for observing the ellipticity spectra $C^\epsilon_{B,i}(\ell)$ (solid lines) and all three positive-parity spectra $C^\epsilon_{E,i}(\ell)$, $C^\epsilon_{S,i}(\ell)$ and $C^\epsilon_{C,i}(\ell)$ combined (dashed lines) as a function of multipole $\ell$ and number of tomographic bins. The sky coverage is set to one. The signal increases with increasing number of tomography bins.}
\label{fig_s2n_comb}
\end{figure}

Similar results can be obtained for a combined measurement of all three positive-parity spectra $C^\epsilon_{E,i}(\ell)$, $C^\epsilon_{S,i}(\ell)$ and $C^\epsilon_{C,i}(\ell)$. Due to the fact that $\epsilon_s$ is derived from the two modes $\epsilon_+$ and $\epsilon_\times$ these measurements are not independent and their covariance needs to be incorporated into the estimate. Fig.~\ref{fig_s2n_comb} shows that this measurement is able to yield $\Sigma_{SEC} = 200\sigma$ ($140\sigma$) when using tomography. The negative-parity spectrum $C^\epsilon_{B,i}(\ell)$, which is predicted to be zero for gravitational lensing, can be easily measured with a significance of about $\Sigma_B = 100\sigma$ (70$\sigma$). Clearly, the lack of cosmic variance due to gravitational lensing makes up for the intrinsically smaller signal.

In summary, we find that ellipticity alignments due to correlated angular momenta should yield a signal which is easily measureable in future tomographic surveys. As the signal is proportional to the alignment parameter $a$ and the disk thickness parameter $\alpha$, one can expect errors on the percent-level on these two parameters of the quadratic alignment model. It should be kept in mind, however, that the product of the parameters determines the amplitude of the intrinsic ellipticity spectrum and that they can not be measured separately. 

From the cosmological parameter set, only $\Omega_m$, $h$, $n_s$ and to a lesser extend $w$ influence the shape of the ellipticity correlation function, while $\sigma_8$ is not relevant in the quadratic model: This can be seen in eqns.~(\ref{eqn_complex}) and~(\ref{eqn_scalar}), where the ellipticity is linked to the traceless tidal shear which does not depend on the amplitude of fluctuations in the tidal shear field as a consequence of the angular momentum model. Secondly, all terms in the ellipticity correlation functions depend on ratios between $\zeta_n$-functions such that the magnitude of $\zeta_n$ is canceled. $\Omega_m$, $h$ and $n_s$ determine the shape of the CDM-spectrum and influence therefore the shape of the ellipticity correlations through the $\zeta_n$-functions. Additionally, the shape of the angular correlations is affected by the conversion between redshift and comoving distance in the Limber-projection, which depends on $\Omega_m$ and $w$.

We conclude that intrinsic alignments are by far the most important secondary effect in weak lensing surveys even at high survey depth, and surpass in particular in $B$-mode generation of other effects such as Born-corrections or clustering. Baryonic effects on the matter spectrum are of similiar order compared to intrinsic alignments.

\section{Parameter estimation}\label{sect_lensing}
Weak lensing shear operates on the complex ellipticity $\epsilon=\epsilon_++\ci\epsilon_\times$ of a galaxy through the mapping $\epsilon\rightarrow\epsilon+\gamma$ in the limit of small shears. Therefore, correlations of the observed ellipticities will contain the intrinsic ellipticity correlation $\bra\epsilon\epsilon^\prime\ket$, the correlation of the lensing shear $\bra\gamma\gamma^\prime\ket$ and possibly cross-correlations between intrinsic shapes and gravitational shear, $\bra\gamma\epsilon^\prime+\gamma^\prime\epsilon\ket$, although the last term is zero in the case of the quadratic alignment model and Gaussian fluctuations. Therefore, it is possible to add lensing and intrinsic alignments as independent contributions to the total ellipticity spectrum.

\subsection{Statistical errors}
The standard method of computing forecasts of statistical errors is the Fisher-matrix technique \citep{1997ApJ...480...22T, 2009arXiv0906.0974B, 2011IJMPD..20.2559B}, in which the cross-correlation between different tomographic bins can be correctly incorporated. We derive forecasts on cosmological parameters from the positive-parity shear spectrum $C^\gamma_{E,ij}(\ell)$ and use the expression
\begin{equation}
F_{\mu\nu} = \frac{f_\mathrm{sky}}{2}\sum_\ell(2\ell+1)\:\trace\left(\partial_\mu\ln C\:\partial_\nu\ln C\right)
\end{equation}
for the Fisher-matrix, where the scaling $\propto f_\mathrm{sky}$ lowers the signal amplitude due to incomplete sky coverage of the survey. In the case of tomographic bins which contain equal fractions of the galaxy sample, the noise to be added to the covariance is diagonal,
\begin{equation}
C_{ij}(\ell) = C^\gamma_{E,ij}(\ell) + \frac{n_\mathrm{bin}}{\bar{n}}\sigma_\epsilon^2\delta_{ij}.
\end{equation}

From the Fisher-matrix it is straightforward to define conditional errors $\sigma_{\mu,c}^2 = 1/F_{\mu\mu}$ and marginal errors $\sigma_{\mu,m}^2=(F^{-1})_{\mu\mu}$ on cosmological parameters, where we use the set $\Omega_m$, $\sigma_8$, $h$, $n_s$ and $w$, while imposing spatial flatness. Commonly, we derive statistical errors by summing over all multipoles until $\ell=3000$, and vary the number of tomography bins between $n_\mathrm{bin}=2$ and $5$, which results in a lensing signal of close to $\Sigma = 1000\sigma$ and marginalised statistical errors on the cosmological parameters on the percent level.

\subsection{Systematic errors}
The presence of intrinsic alignments in weak lensing data would lead to biases in the estimation of cosmological parameters if they are uncorrected. For this case we consider biases that would arise if an alignment contribution $C^\epsilon_{E,i}(\ell)$ is added to the weak lensing spectra $C^\gamma_{E,ii}(\ell)$ for equal bin indices, while $C^\gamma_{E,ij}(\ell)$, $i\neq j$, is unchanged. A fit of $C^\gamma_{E,ij}(\ell)$ to the combined $C^\gamma_{E,ij}(\ell) + C^\epsilon_{E,i}(\ell)\delta_{ij}$ (no summation over $i$ implied) would then give rise to a parameter estimation bias $\delta_\mu$. These estimation biases can be computed from the true model $C_{t,ij}=C^\gamma_{E,ij}+C^\epsilon_{E,i}\delta_{ij}$ and the false, incomplete model $C_{f,ij}=C^\gamma_{E,ij}$ by solving the linear equation
\begin{equation}
\sum_\nu G_{\mu\nu}\delta_\nu = a_\mu\rightarrow \delta_\mu = \sum_\nu (G^{-1})_{\mu\nu}a_\nu,
\end{equation}
with the vector $a_\mu$,
\begin{equation}
a_\mu = \sum_\ell\frac{2\ell+1}{2}
\trace\left[\partial x_\mu\ln C_f\left(\id-C_f^{-1}C_t\right)\right],
\end{equation}
and the matrix $G_{\mu\nu}$,
\begin{eqnarray}
G_{\mu\nu}
& = & \sum_\ell\frac{2\ell+1}{2}
\trace
\left[C_f^{-1}\:\partial^2_{\mu\nu}C_f\:\left(C_f^{-1}C_t-\id\right)\right]
\nonumber\\
& - & \sum_\ell\frac{2\ell+1}{2}
\trace
\left[\partial_\mu\ln C_f\partial_\nu\ln C_f\:\left(2C_f^{-1}C_t-\id\right)\right],
\end{eqnarray}
where $\id$ refers to the unit matrix in $n_\mathrm{bin}$ dimensions. The formalism employed here \citep{2012MNRAS.423.3445S} is a generalisation of the bias estimation formalism proposed by \citet{2007MNRAS.381.1347C}, \citet{2009MNRAS.392.1153T}, \citet{2008MNRAS.391..228A} and \citet{2011MNRAS.tmp..612M} to tomographic data. It has been shown by comparison to the results from Monte-Carlo Markov-chains that it predicts estimation biases well, at least in the case of small systematics \citep{2010MNRAS.404.1197T}.

\begin{figure}
\begin{center}
\resizebox{\hsize}{!}{\includegraphics{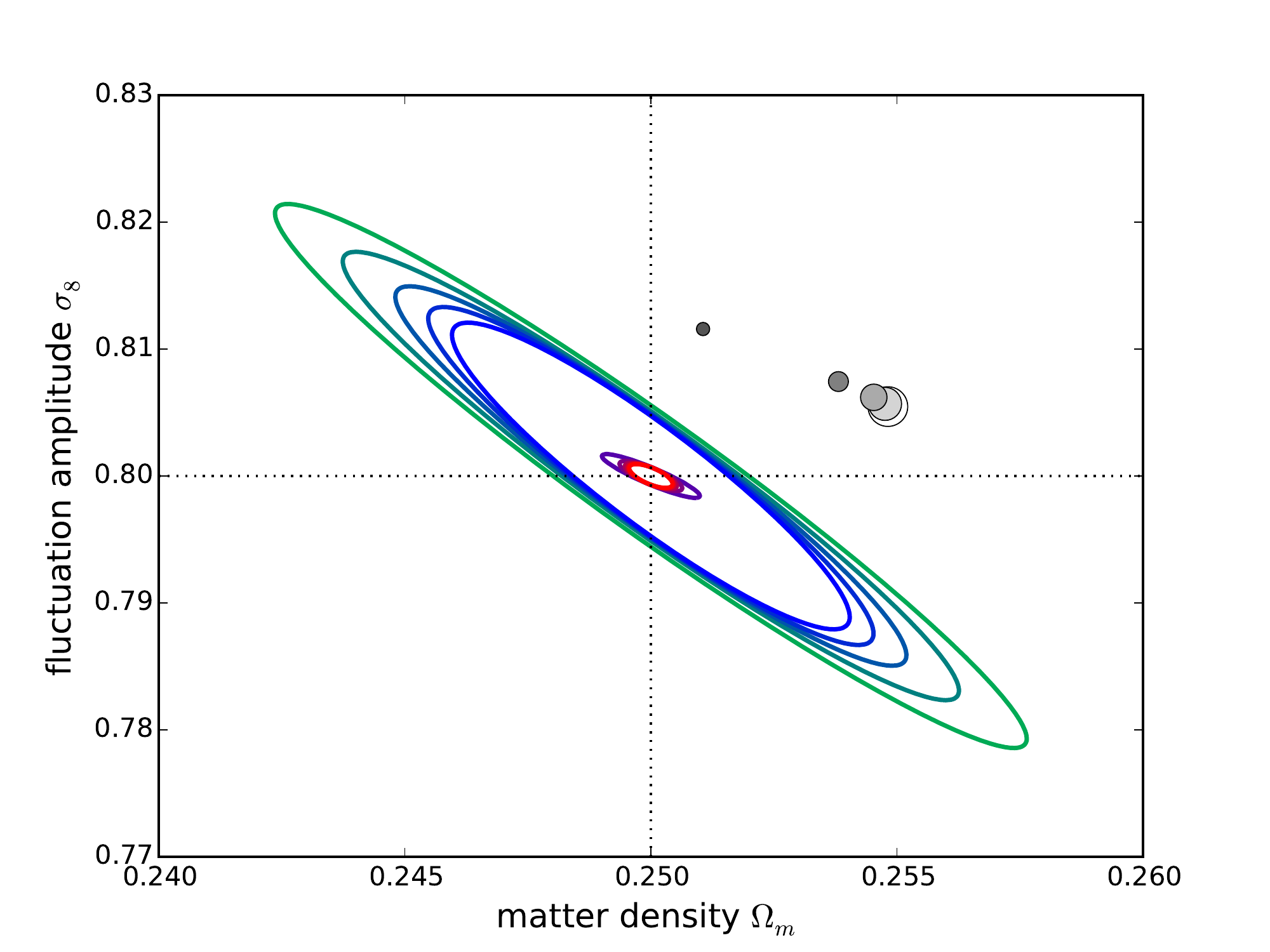}}
\end{center}
\caption{Marginalised likelihood contours (blue-green) and conditional likelihood contours (red-blue) at $1\sigma$-confidence for the parameter pair $\Omega_m$ and $\sigma_8$, for $n_\mathrm{bin}=2$ (smallest) up to $6$ (largest), in comparison to parameter estimation biases due to intrinsic alignments with $a=0.25$, all for $\ell_\mathrm{max}=3000$.}
\label{fig_bias_os}
\end{figure}

Fig.~\ref{fig_bias_os} shows parameter estimation biases in comparison to $1\sigma$-likelihood contours on $\Omega_m$ and $\sigma_8$, both marginalised over the entire parameter set and assuming that all other parameters are perfectly known. Intrinsic alignments bias both $\sigma_8$ and $\Omega_m$ high by a few percent, which should be see in comparison to the very good statistical precision. The physical reason for biasing is the addition of power on the spectrum by intrinsic alignments, which needs to be compensated by a fit through increasing the amplitude of the signal, leading to larger values of both parameters. Another interesting trend is a growing bias in the cosmological parameters when the number of tomography bins is increased.

\begin{figure}
\begin{center}
\resizebox{\hsize}{!}{\includegraphics{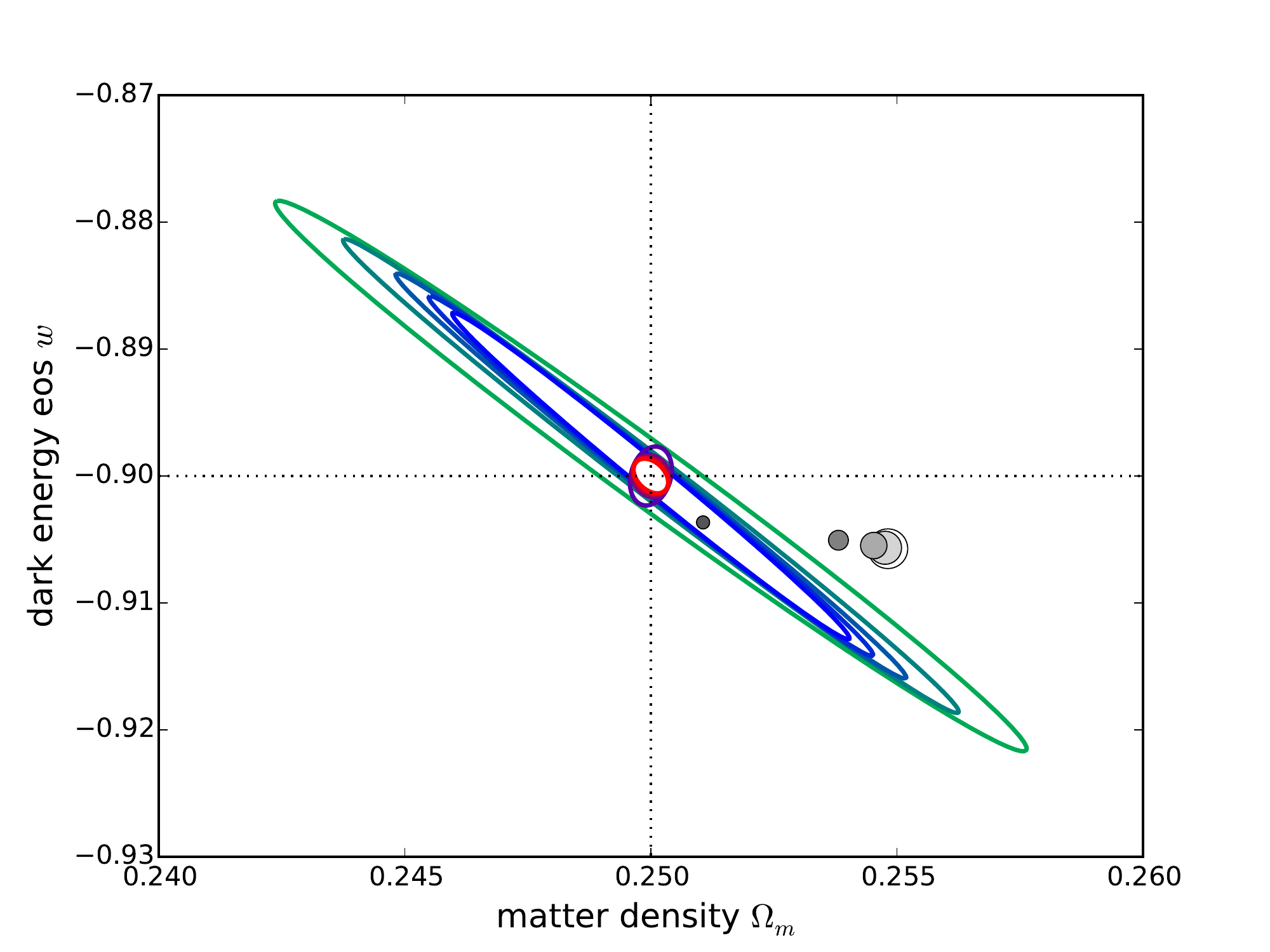}}
\end{center}
\caption{Marginalised (blue-green) and conditional (red-blue) $1\sigma$-contours in the $\Omega_m$-$w$-plane, along with parameter estimation biases. The colour indicates the number of tomography bins used in the analysis, ranging from $n_\mathrm{bin}=2$ (smallest) to $6$ (largest), for $\ell_\mathrm{max}=3000$.}
\label{fig_bias_ow}
\end{figure}

Fig.~\ref{fig_bias_ow} repeats the above analysis for the cosmological parameters $\Omega_m$ and $w$ by showing the biases in relation to marginalised and conditional likelihood contours. In particular $w$ is measured more negative than the fiducial model, hence one could potentially mistake a dark energy model for $\Lambda$CDM, again because the intrinsic alignments increase the amplitude of the ellipticity spectrum. Again, one finds increasing biases if the lensing survey is divided into more redshift bins.

\begin{figure}
\begin{center}
\resizebox{\hsize}{!}{\includegraphics{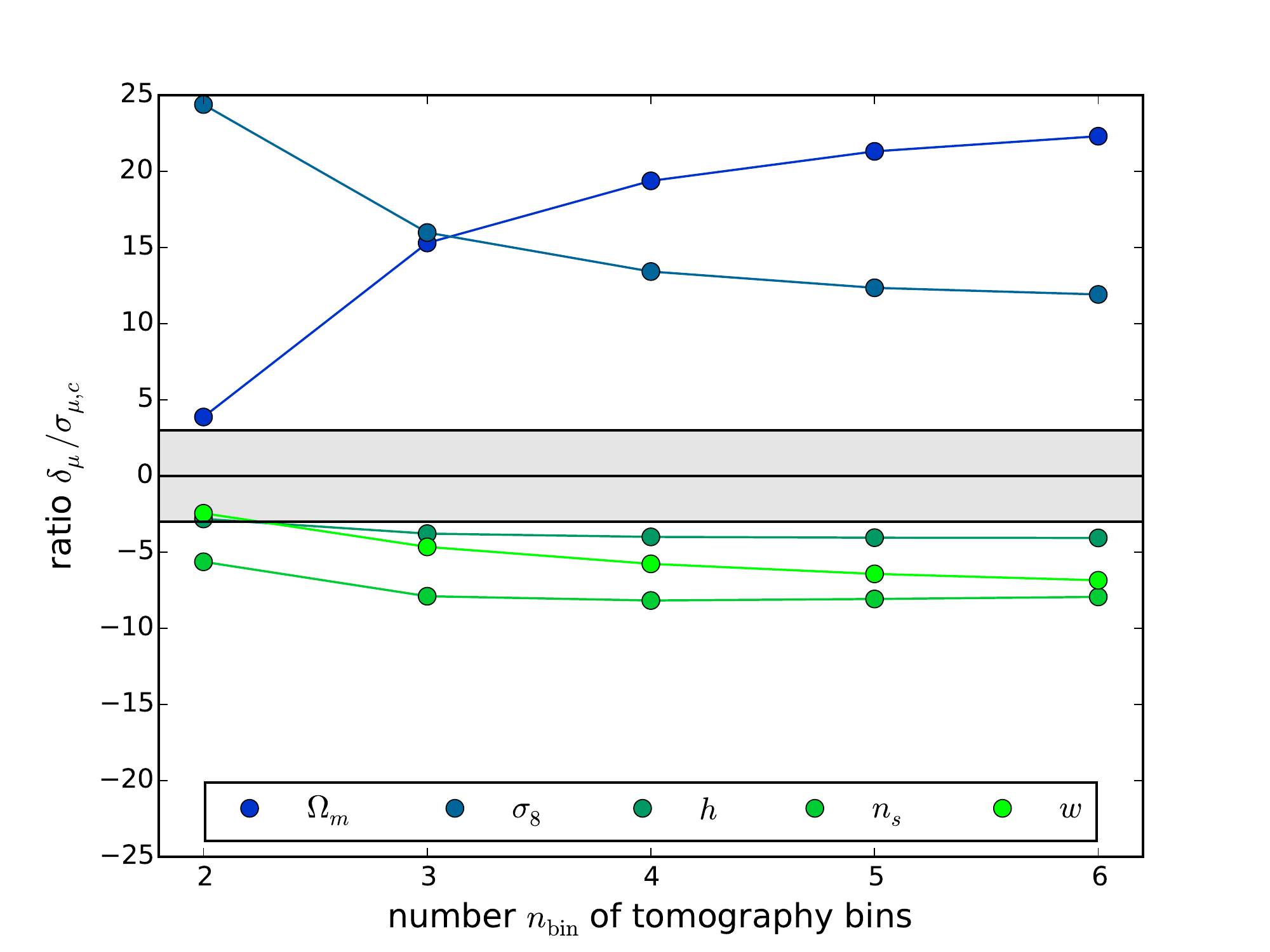}}
\end{center}
\caption{Systematic error $\delta_\mu$ in units of the conditional statistical error $\sigma_{\mu,c}$ in a $w$CDM parameter set consisting of $\Omega_m$, $\sigma_8$, $h$, $n_s$ and $w$, as a function of tomography bins $n_\mathrm{bin}$ and for $\ell_\mathrm{max}=3000$. The shaded area indicates the $3\sigma$-interval.}
\label{fig_ratio}
\end{figure}

\begin{figure}
\begin{center}
\resizebox{\hsize}{!}{\includegraphics{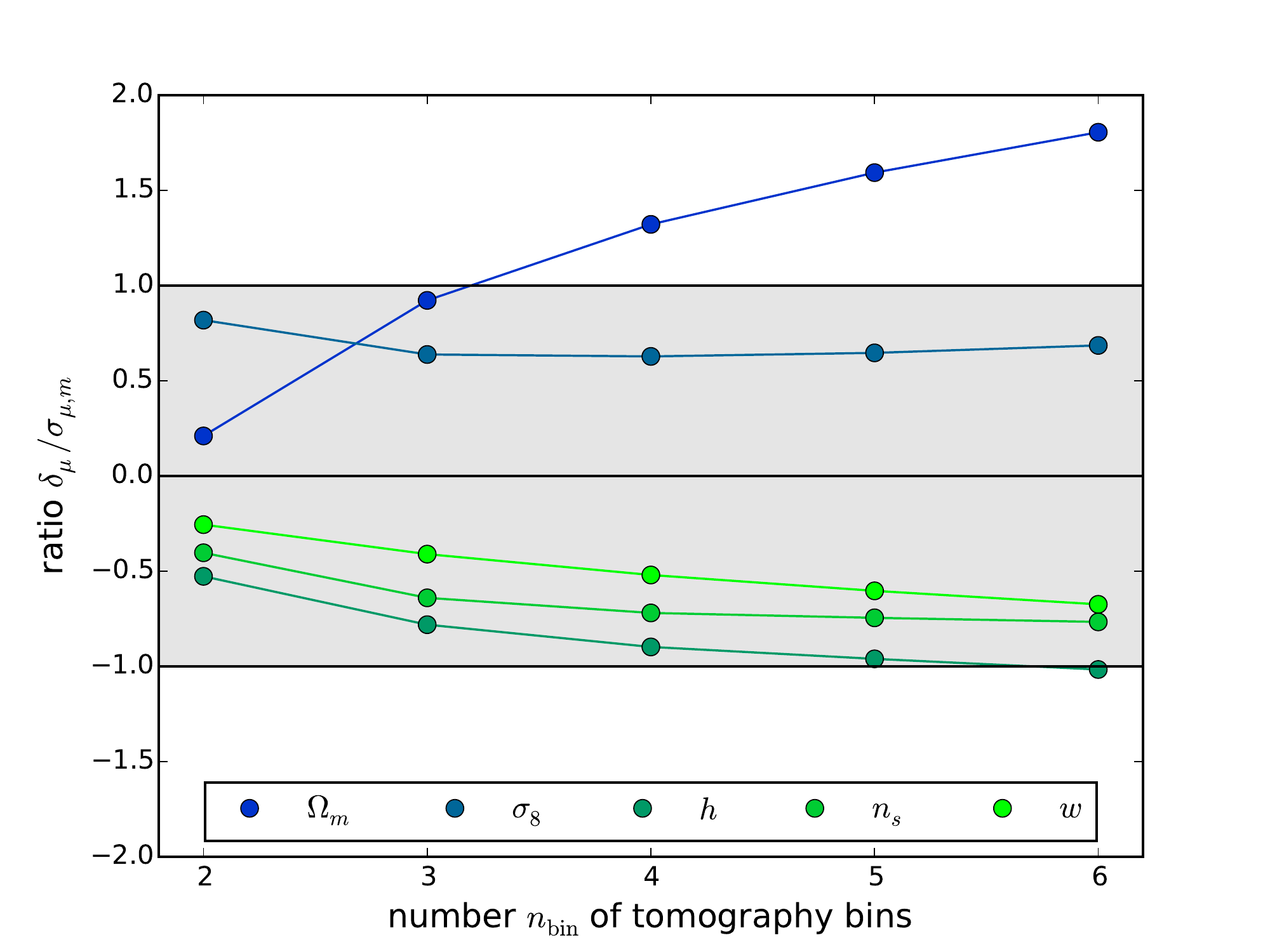}}
\end{center}
\caption{Systematic error $\delta_\mu$ in units of the marginalised statistical error $\sigma_{\mu,m}$ in a $w$CDM parameter set consisting of $\Omega_m$, $\sigma_8$, $h$, $n_s$ and $w$, as a function tomography bins $n_\mathrm{bin}$ and $\ell_\mathrm{max}=3000$. The shaded area indicates the $1\sigma$-interval.}
\label{fig_ratio_marginalise}
\end{figure}

These results are presented by Figs.~\ref{fig_ratio} and~\ref{fig_ratio_marginalise} in a more quantitative way, by illustrating the dependence of the ratio $\delta_\mu/\sigma_\mu$ for the conditional and marginalised statistical error, respectively.

Naturally, the biases $\delta_\mu/\sigma_{\mu,c}$ are much larger than $\delta_\mu/\sigma_{\mu,m}$ because statistical uncertainties are smaller in models with fewer parameters. But in general, the entire $\Lambda$CDM- or $w$CDM-parameter set is affected by intrinsic alignments. Which parameters suffer most depends on the number of tomography bins being chosen. With 5 bins and $\ell_\mathrm{max}=3000$ it is $\Omega_m$, $\sigma_8$ and $n_s$ that are most strongly affected, followed by the dark energy equation of state $w$.

Comparing the bias to either the marginalised or conditional error in projections, as done by Figs.~\ref{fig_bias_os} and~\ref{fig_bias_ow}, does not give a fair representation of biasing relative to the magnitude of the statistical errors and their degeneracies. For that reason, we define the figure of bias $q$ as a quadratic form,
\begin{equation}
q^{2} = \sum_{\mu\nu}\delta_\mu F_{\mu\nu}\delta_\nu
\label{eqn_alpha}
\end{equation}
which compares the systematical error $\delta_\mu$ with the statistical one and reflects the orientation of the vector $\delta_\mu$ with the shape and size of the Fisher-ellipse in the full parameter space. In the limit of absent degeneracies, $q$ corresponds to the quadratic sum of the bias components normalised by the statistical errors on individual parameters. At the same time, $\exp(-q^2/2)$ is the evalutated likelihood at the position $\delta_\mu$ relative to the fiducial model.

We think that in this way $q$ is a convenient way of quantifying the total amount of estimation bias relative to the total statistical uncertainty of a measurement, but has the disadvantage of not containing any information about the biasing direction as eqn.~(\ref{eqn_alpha}) is a positive definite quadratic form. Values of $q$ are typically $\simeq 20$ for $\ell=3000$ with a weak variation with the number of bins, so rather in accordance with the ratio $\delta_\mu/\sigma_{\mu,c}$, where the bias was expressed in units of the conditional statistical error, than with $\delta_\mu/\sigma_{\mu,m}$, with the marginalised statistical error.

\section{Summary}\label{sect_summary}
Subject of this paper are angular spectra of angular-momentum induced, quadratic alignments of galaxies as they would appear in tomographic weak lensing surveys, in particular in that of a Euclid-like survey. Starting from tidal torquing as a model for generating galaxy spin we derive the statistics of galaxy ellipticities for a given tidal shear and compute the ellipticity correlation function resulting from the tidal shear correlations. Projection and subsequent Fourier-transform yield the angular ellipticity correlation function and ultimately the angular ellipticity spectrum, whose statistical properties we analyse in detail.

Very important simplifications in our analysis include the prediction of the angular momentum direction by linear tidal torquing and the description of the angular momentum field to be derived from the tidal shear field in a Gaussian, local random process. Furthermore, we assume that the baryonic component aligns itself perfectly with the angular momentum axis of the host halo: This, in particular, is a serious simplification which we need to impose in our analytical work, such that the derived signal to noise-ratios should be considered as upper limits.
\begin{enumerate}
\item{Ellipticity correlation functions show that the ellipticity field is only correlated on scales smaller than a few arcminutes due to angular-momentum induced alignments, and the ellipticity correlation quickly vanishes on scales larger than ten arcminuts for all components of the ellipticity.}
\item{Angular ellipticity spectra follow from ellipticity correlation functions through Fourier-transform. We isolate the four non-zero spectra: the $E$-mode spectrum $C^\epsilon_E(\ell)$, the associated $B$-mode spectrum $C^\epsilon_B(\ell)$, the spectrum $C^\epsilon_S(\ell)$ of the ellipticity modulus and finally the cross-spectrum $C^\epsilon_C(\ell)$ of the $E$-mode and the ellipticity modulus. These four spectra correspond formally to the four non-zero temperature and polarisation spectra of the cosmic microwave background. Ellipticity spectra are dominated by weak gravitational lensing on small multipoles, but have similar amplitudes on multipoles of a few hundred.}
\item{The significance at which the spectra can be measured was estimated assuming Gaussianity of the ellipticity and the weak lensing fields. Euclid turns out to be an excellent mission for investigating intrinsic alignments: Due to the fact that intrinsic alignments are effectively uncorrelated over large distances, each redshift bin provides a statistically independent estimate of the correlation function, only with larger Poissonian shape noise. Individual angular ellipticity spectra can be measured at significances for the spectrum $C^\epsilon_E(\ell)$ of $\sim10\sigma$ for a non-tomographic survey increasing up to values in excess of $100\sigma$ if the survey is divided into 9 redshift bins. Similarly, the ellipticity modulus correlation can be measured with $\sim6\sigma$, and this number can be boosted to $\sim60\sigma$ by doing 9-bin-tomography. Corresponding numbers for the $B$-mode spectrum $C^\epsilon_C(\ell)$ are $\sim20\sigma$ and $\sim90\sigma$, respectively. To put these significances in context, we would like to emphasise that tomographic weak lensing generates a signal close to $1000\sigma$.}
\item{While the four angular ellipticity spectra $C^\epsilon_E(\ell)$, $C^\epsilon_B(\ell)$, $C^\epsilon_C(\ell)$ and $C^\epsilon_S(\ell)$ are not statistically independent because they are derived from only two components of the ellipticity field, the combination of all positive-parity spectra would yield a total signal of almost $200\sigma$ using 9 tomographic bins and extending the estimates to multipoles of $\ell=3000$. Due to the fact that signal to noise-ratios are inversely proportional to the relative error on the normalisation of the signal, we could expect percent-level errors on the alignment parameters $\alpha$ and $a$.}
\item{Estimates of cosmological parameters from weak gravitational lensing are severely biased by intrinsic alignments. While the absolute values of the biases do not change much when increasing the number of tomography bins, the bias in units of the statistical error increases significantly: We find biases of $+5\sigma_w$ for the dark energy equation of state parameter $w$, $-20\sigma_{\Omega_m}$ for the matter density $\Omega_m$ and $-12\sigma_{\sigma_8}$ for the spectrum normalisation $\sigma_8$. These biases have important implications for measurements of parameters as well as for the selection of cosmological models, as one might prefer $\Lambda$CDM over a true dark energy model.}
\end{enumerate}

In future work we aim at supplementing the angular-momentum based alignment model, which is thought to be applicable to spiral galaxies, with a tidal shearing model for treating elliptical galaxies \citep{HS04,HS10,BMS+12,BMS11,2015arXiv150402510B}, and combine both models for a realistic morphological mix of galaxy types. This would have important consequences: Due to the fact that there should be no intrinsic cross-correlation between the shapes of spiral and elliptical galaxies for Gaussian random fields one can expect smaller levels of intrinsic alignments. If $q$ is the fraction of spiral galaxies in a sample, the total ellipticity correlation function would be $\bra\epsilon\epsilon^\prime\ket = q^2\bra\epsilon_s\epsilon_s^\prime\ket + (1-q)^2\bra\epsilon_e\epsilon_e^\prime\ket$ with a correlation function $\bra\epsilon_s\epsilon_s^\prime\ket$ of the shapes of spiral galaxies and $\bra\epsilon_e\epsilon_e^\prime\ket$ of elliptical galaxies, respectively. Both factors $q^2$ and $(1-q)^2$ would be smaller than one, thus reducing the amount of ellipticity correlation relative to the assumption of a single alignment model. Conversely, intrinsic alignments have been observationally confirmed for elliptical galaxies \citep[for instance by CFHTLenS, ][]{KAH+14,fu14}, and those have been demonstrated to have an impact on parameter estimation, in particular in models with modified gravity \citep{LBK+12, DIP+15}. The impact of alignments of elliptical galaxies on parameter masurement from weak lensing has been estimated by \citet{JMA+11} for the Mega-Z sample, and biases were found to be significant, even for a rather small data set. 

A second interesting topic are cross-correlations between the shapes of elliptical galaxies and weak gravitational lensing, which are absent in the case of spiral galaxies, due to the same argument as before: Lensing, as the shapes of elliptical galaxies, are linear in the tidal shear. These cross-correlations are negative and shape the ellipticity spectrum in a more complicated way with consequences for biases on cosmological parameters. Furthermore, both types of intrinsic alignments are local and should not show correlations across different tomography bins, and only quadratic alignments should be able to excite $B$-modes at lowest order. All these properties would be important validations of the basic ideas behind galaxy alignments.

\section*{Acknowledgements}
We would like to thank Luca Amendola, Vanessa B{\"o}hm, Jens Jasche, Benjamin Joachimi and Tom Kitching for valuable comments, and Vanessa B{\"o}hm in particular for providing the value of the misalignment parameter $a$ from her simulations.

\bibliography{bibtex/aamnem,bibtex/references}
\bibliographystyle{mn2e}

\appendix

\bsp

\label{lastpage}

\end{document}